\documentclass[11pt]{article}
\usepackage{graphicx}
\usepackage{url}
\usepackage{rotating}
\usepackage{bm}
\usepackage{amsmath}
\usepackage{relsize}
\usepackage{multirow}
\usepackage{natbib}

\def\be{\begin{equation}}
\def\ee{\end{equation}}

\def\be{\begin{equation}}
\def\bea{\begin{eqnarray}}
\def\ee{\end{equation}}
\def\eea{\end{eqnarray}}

\def\R0{R$_{\odot}$}

\def\pccm6{pc cm$^{-6}$}

\def\h2o{H$_2$O}
\def\bs{\boldsymbol}

\newcommand{\threevdots}{%
  \vbox{\baselineskip1ex\lineskiplimit0pt%
  \hbox{.}\hbox{.}\hbox{.}}}

\begin{document}

\begin{titlepage}
\begin{center}

\textsc{\LARGE \bf NATIONAL RADIO ASTRONOMY OBSERVATORY\\[0.4cm] Charlottesville, Virginia}\\[5cm]

\textsc{\Large \bf ELECTRONICS DIVISION INTERNAL REPORT NO. 331}\\[5cm]

{\LARGE \bfseries Phased Array Feed Model Equations corresponding to two definitions of 
embedded beam pattern \\[1.0cm] }

{\Large
D. Anish Roshi$^{1}$ \\ [0.4cm]
{\small
$^{1}$ National Radio Astronomy Observatory, Charlottesville. \\
}
}

\vfill
September 28, 2017

\end{center}
\end{titlepage}

\title{Phased Array Feed Model Equations corresponding to two definitions of embedded beam pattern}
\author{D. Anish Roshi}
\date{September 28, 2017 \\Version 0.2}
\maketitle

\section*{Abstract}
In this report, we present the phased array feed (PAF) model equations
for two definitions of embedded beam patterns. In Roshi \& Fisher (2016), 
we presented the PAF model by defining the embedded beam pattern
as the beam pattern due to a 1 V excitation to one port and all other
ports short circuited. This embedded beam pattern is referred to as
voltage-embedded-beam (VEB). The embedded beam pattern can also be defined as
the beam pattern due to a 1 A excitation to one port and all other
ports open circuited. This definition is usually used in engineering
literature and we refer to the pattern as current-embedded-beam (CEB). 
Here we derive the relationship between the two embedded
beam patterns and present the corresponding model equations.

\section{Introduction}

\subsection{Embedded beam pattern}

\begin{figure}[t]
\begin{tabular}{cc}
\includegraphics[width=3.0in, height=2.5in, angle =0]{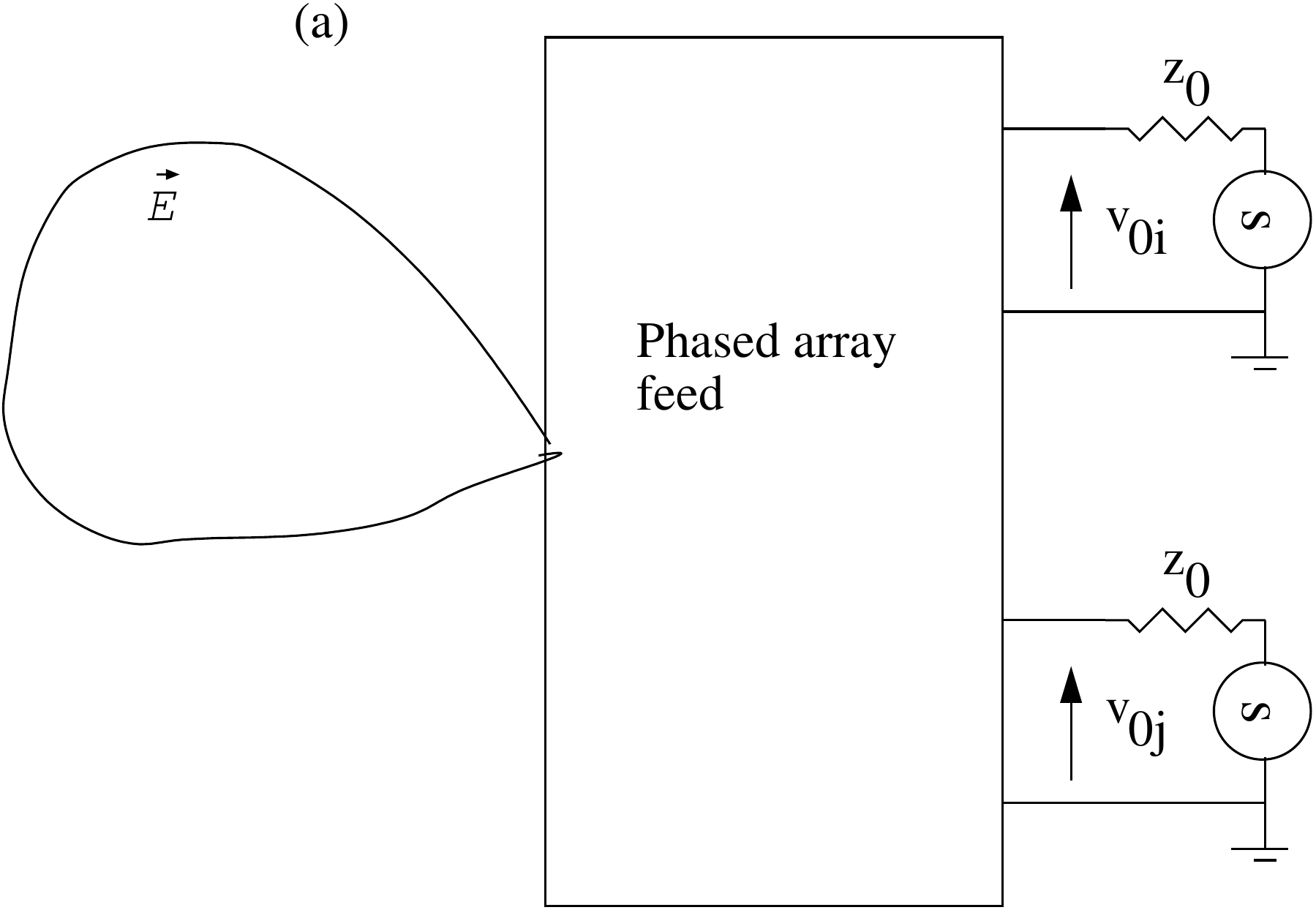} & 
\includegraphics[width=3.0in, height=2.5in, angle =0]{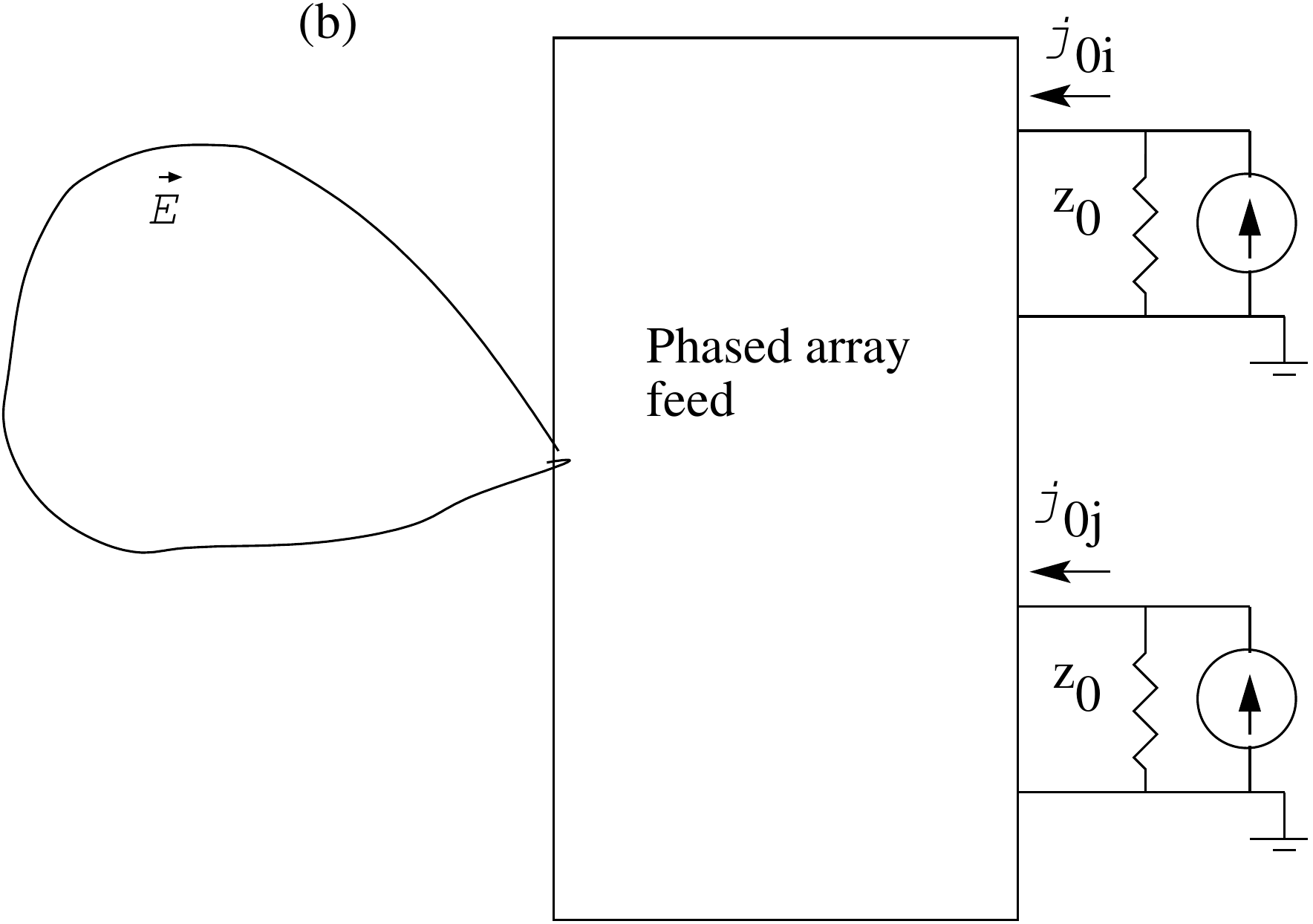} 
\end{tabular} 
\caption{PAF in transmitting mode. The net radiation pattern of the PAF
can be expressed as a `weighted' sum of the embedded beam patterns (see
Eq.~\ref{Evsvembpat} \& \ref{Evsiembpat}). The `weights' can be either the
port voltages along with the VEB, 
$\bs{\vec{\mathcal{E}}^e}$ (left)
or port currents along with the CEB,  
$\bs{\vec{\mathcal{\psi}}^e}$ (right).
}
\label{fig1}
\end{figure}

It is convenient to express the radiation pattern of PAF in terms of the {\em embedded beam
pattern} (see Roshi \& Fisher 2016). The embedded beam pattern can be defined 
in different ways. A definition 
used in Roshi \& Fisher (2016) is : the $j^{th}$ embedded beam pattern, $\vec{\mathcal{E}}^e_j$ is 
the beam pattern of the PAF when $j^{th}$ port is excited with
1 V (i.e. $v_{0_j} = 1$ V) and all other ports are short circuited 
(i.e. $v_{0_i} = 0$ V for $i \neq j$; see Fig.~\ref{fig1}a).
The source impedance for excitation is considered to be equal to $z_0$, the 
characteristic impedance of the transmission line connected to the dipole.
There will be $M$ embedded beam patterns for a $M$ element PAF, which 
are represented conveniently as a vector $\bs{\vec{\mathcal{E}}^e}$,
\be
\bs{\vec{\mathcal{E}}^e}^T = \left[\vec{\mathcal{E}}^e_1, \vec{\mathcal{E}}^e_2, ... \right].
\ee 
These beam patterns are functions of the position vector $\vec{r}$ with the origin of the
coordinate system located at the center of the PAF (see Roshi \& Fisher 2016). 
The beam patterns are specified at the far-field 
ie $|\vec{r}| >> \frac{2 D_{array}^2}{\lambda}$, where $D_{array}$ is the maximum
physical size of the PAF and $\lambda$ is the wavelength of operation of the PAF.  
The radiation pattern when the PAF is excited by an arbitrary set of port voltages
is obtained by scaling the embedded beam patterns with the
port voltage and summing them up.
Hence the dimension of the embedded beam pattern in this definition is m$^{-1}$. 
At the far-field, the beam pattern can be described by an outgoing spherical wave,
\be
\vec{\mathcal{E}}^e_i(\vec{r}) =  \vec{E}^e_i(\theta, \phi) \; \frac{e^{j\vec{k}.\vec{r}}}{r},
\label{farf}
\ee
where $\vec{\mathcal{E}}^e_i$ is the $i^{th}$ embedded beam pattern,
$r$ and $\hat{r}$ are the magnitude and the unit vector in the direction 
of $\vec{r}$ respectively, $\vec{k} = \frac{2\pi}{\lambda} \hat{r}$ is 
the propagation vector. Here $\vec{E}^e_i$ depends only 
on the coordinates $\theta, \phi$. The geometric phase due to the location
of elements (or in other words the excitation current distribution) away
from the co-ordinate center is included in $\vec{E}^e_i$. From the 
definition of embedded pattern it follows that $\vec{E}^e_i$ is dimensionless. 
The fields here are harmonic quantities, and for simplicity we omit the term $e^{j\omega t}$.
The radiation pattern of the PAF when excited by a set of arbitrary
port voltages is 
\bea 
\vec{\mathcal{E}}(\vec{r}) & =  & \sum_{i=1,M} v_{0_i} \vec{\mathcal{E}}^e_i(\vec{r}), \nonumber \\ 
                           & =  & \bm V_0^T \bs{\vec{\mathcal{E}}^e},  
\label{PAFfpat}
\eea
where $\bm V_0$ is the vector of port voltages $v_{0_i}$ (see Fig.~\ref{fig1}a). The 
radiation pattern $\vec{\mathcal{E}}$ has units V/m. In the far-field,
the ($\theta, \phi$) dependence of the radiation pattern can be written
in a similar fashion,
\be
\vec{E}(\theta, \phi) = \bm V_0^T \bs{\vec{E}^e}.
\label{Evsvembpat}
\ee
The unit of $\vec{E}$ is V. In this report, we refer to both $\bs{\vec{\mathcal{E}}^e}$ 
and $\bs{\vec{E}^e}$ as embedded beam pattern VEB. 

Another definition for embedded beam pattern is: the $j^{th}$ embedded beam pattern, 
$\vec{\mathcal{\psi}}^e_j$ is 
the beam pattern of the PAF when $j^{th}$ port is excited with
1 A and all other ports are open circuited, i.e.
\bea
\mathcal{J}_{0_i} & = & 1\;\; \textrm{A}\;\; \textrm{for}\; i = j,  \nonumber\\
\mathcal{J}_{0_i} & = & 0\;\; \textrm{A}\;\; \textrm{for}\; i \neq j,
\eea
where $\mathcal{J}_{0_i}$ are the port currents.
The source impedance for excitation is considered to be equal to $z_0$.  
As before there are $M$ embedded beam patterns, which are represented conveniently
as a vector $\bs{\vec{\mathcal{\psi}}^e}$,
\be
\bs{\vec{\mathcal{\psi}}^e}^T = \left[\vec{\mathcal{\psi}}^e_1, \vec{\mathcal{\psi}}^e_2, ... \right].
\ee 
The beam pattern at far field can be written as,
\be
\vec{\mathcal{\psi}}^e_i(\vec{r}) =  \vec{\Psi}^e_i(\theta, \phi) \; \frac{e^{j\vec{k}.\vec{r}}}{r},
\label{ifarf}
\ee
The radiation pattern of the PAF when excited by a set of arbitrary
port currents is 
\bea 
\vec{\mathcal{E}}(\vec{r}) & =  & \sum_{i=1,M} \mathcal{J}_{0_i} \vec{\mathcal{\psi}}^e_i(\vec{r}), \nonumber \\ 
                           & =  & \bm I_0^T \bs{\vec{\mathcal{\psi}}^e},  
\label{PAFfpat2}
\eea
where $\bm I_0$ is the vector of port currents $\mathcal{J}_{0_i}$ (see Fig.~\ref{fig1}b). The 
radiation pattern $\vec{\mathcal{E}}$ has the unit V/m and
$\vec{\mathcal{\psi}}^e_i$ has unit V/A/m. As before, 
the ($\theta, \phi$) dependence of the far-field radiation pattern can be written
as,
\be
\vec{E}(\theta, \phi) = \bm I_0^T \bs{\vec{\Psi}^e}.
\label{Evsiembpat}
\ee
The unit of $\vec{E}$ is V and that of $\vec{\Psi}^e$ is V/A. In this report,
we refer to both $\bs{\vec{\mathcal{\psi}}^e}$ and $\bs{\vec{\Psi}^e}$ as the 
embedded beam pattern CEB. 

The relationship between the two embedded beam patterns can be obtained
using the network relationship between the port voltages and
currents, $\bm V_0 = \bm Z \bm I_0$.
Substituting this relationship in Eq.~\ref{PAFfpat}, we get 
\be
\vec{\mathcal{E}}  =  \bm V_0^T \bs{\vec{\mathcal{E}}^e} =  \bm I_0^T \bm Z^T \bs{\vec{\mathcal{E}}^e}.
\label{eq11}
\ee
From Eq.~\ref{PAFfpat2} \& \ref{eq11} it follows
\be
\bs{\vec{\mathcal{\psi}}^e} = \bm Z^T \bs{\vec{\mathcal{E}}^e}
\ee
For a reciprocal PAF, $\bm Z^T = \bm Z$, and so the above equation 
can also be written as
\be
\bs{\vec{\mathcal{\psi}}^e} = \bm Z \bs{\vec{\mathcal{E}}^e}
\ee

\section{PAF model equations corresponding to the two embedded beam patterns}

The PAF model equations are somewhat simplified when written in terms  
of CEB $\bs{\vec{\mathcal{\psi}}^e}$. Essentially in almost 
all relevant equations the impedance matrix $\bm Z$ is absorbed in the
embedded beam pattern when $\bs{\vec{\mathcal{\psi}}^e}$ is used. For example,
the open circuit voltage vector (see Eq. 37 in Roshi \& Fisher 2016)
at the output of the PAF for VEB, 
$\bs{\vec{\mathcal{E}}^e}$, and CEB, $\bs{\vec{\mathcal{\psi}}^e}$, is given by Eqs.~\ref{eq14a} \& \ref{eq14b}
respectively; 
\bea
\bm V_{oc} &  = & \bm Z \int_{A_{free}} \left(\bs{\vec{\mathcal{E}}^e}^T
             \times \bs{\mathcal{I}} \vec{\mathcal{H}_r} -
             \bs{\mathcal{I}} \vec{\mathcal{E}_r} \times \bs{\vec{\mathcal{H}}^e}\right)
             \cdot \hat{n}\; \textrm{d}A, \label{eq14a} \\
           &  = & \int_{A_{free}} \left(\bs{\vec{\mathcal{\psi}}^e}^T
             \times \bs{\mathcal{I}} \vec{\mathcal{H}_r} -
             \bs{\mathcal{I}} \vec{\mathcal{E}_r} \times \bs{\vec{\mathcal{J}}^e}\right)
             \cdot \hat{n}\; \textrm{d}A. \label{eq14b}
\label{voc}
\eea 
Here $\bs{\vec{\mathcal{H}}^e}$ and $\bs{\vec{\mathcal{J}}^e}$ are the magnetic 
field patterns corresponding to the VEB, $\bs{\vec{\mathcal{E}}^e}$
and the CEB, $\bs{\vec{\mathcal{\psi}}^e}$ respectively, $\vec{\mathcal{E}_r}$ and
$\vec{\mathcal{H}_r}$ are the incident electric and magnetic fields on the PAF respectively, 
$\bs{\mathcal{I}}$ is the identify matrix, and the integration is over a region outside the
PAF (see Fig. 2 in Roshi \& Fisher 2016). The notation used
in Eq.~\ref{voc} is explained in Appendix J of Roshi \& Fisher (2016).
A list of model equations corresponding to the VEB, 
$\bs{\vec{\mathcal{E}}^e}$ (left) and the CEB, $\bs{\vec{\mathcal{\psi}}^e}$ (right) is given below.
\begin{align}
\bm R_{spill} & =  \frac{4 k_B T_g}{z_f} \bm Z \bm C_{Ce1}, \bm Z^H  
&\bm R_{spill}  = & \frac{4 k_B T_g}{z_f} \bm C_{C\psi1},  \\
\bm R_{signal}& =  \frac{2 S_{source}}{z_f} \bm Z \bm C_{Ie}, \bm Z^H
&\bm R_{signal} = & \frac{2 S_{source}}{z_f} \bm C_{I\psi},  \\
T_{spill}     & =  T_g
                  \frac{\bm w_1^H \bm Z \bm C_{Ce1} \bm Z^H  \bm w_1}
                  {\bm w_1^H \bm Z \bm C_{Ce} \bm Z^H  \bm w_1},  
&T_{spill}      = &  T_g
                  \frac{\bm w_1^H \bm C_{C\psi1}  \bm w_1}
                  {\bm w_1^H \bm C_{C\psi}  \bm w_1},  \\
T_A           & =  \frac{S_{source}}{2 k_B} \frac{\bm w_1^H \bm Z \bm C_{Ie} \bm Z^H \bm w_1}
                  {\bm w_1^H \bm Z \bm C_{Ce} \bm Z^H   \bm w_1},
&T_A            = & \frac{S_{source}}{2 k_B} \frac{\bm w_1^H \bm C_{I\psi} \bm w_1}
                  {\bm w_1^H  \bm C_{C\psi}  \bm w_1} ,\\
\eta_{app}      & =  \frac{1}{A_{ap}} \; \; \frac{\bm w_1^H \bm Z \bm C_{Ie} \bm Z^H \bm w_1}
                  { \bm w_1^H \bm Z \bm C_{Ce} \bm Z^H \bm w_1} ,
&\eta_{app}      = & \frac{1}{A_{ap}} \; \; \frac{\bm w_1^H \bm C_{I\psi} \bm w_1}
                  { \bm w_1^H \bm C_{C\psi} \bm w_1}. 
\end{align}
Here $\bm R_{spill}$, $\bm R_{signal}$ are the open circuit voltage correlations due to spillover
noise and that due to radiation from source respectively, $T_{spill}$ is the spillover
temperature and $T_A$ is the antenna temperature due to the source, $\eta_{app}$ is the
aperture efficiency, $\bm w_1$ is the weight vector applied on the open circuit voltage correlations
(see Roshi \& Fisher 2016),  
\bea
\bm C_{Ce1} & \equiv & \int_{\Omega_{spill}} \bs{\vec{E^e}}\cdot \bs{\vec{E^e}}^H \textrm{d}\Omega,
\label{eqce1} \\
\bm C_{C\psi1} & \equiv & \int_{\Omega_{spill}} \bs{\vec{\Psi}^e}\cdot \bs{\vec{\Psi}^e}^H \textrm{d}\Omega, \label{eqcsi1}\\
\bm C_{Ce} & \equiv & \int_{4\pi} \bs{\vec{E^e}}\cdot \bs{\vec{E^e}}^H \textrm{d}\Omega, \\
\bm C_{C\psi} & \equiv & \int_{4\pi} \bs{\vec{\Psi}^e}\cdot \bs{\vec{\Psi}^e}^H \textrm{d}\Omega,\\
\bm C_{Ie} & \equiv & \left(\int_{A_{pap}} \bs{\vec{\mathcal{E}}^e_{pap}} \textrm{d} A \right) \cdot
                  \left(\int_{A_{pap}} \bs{\vec{\mathcal{E}}^e_{pap}} \textrm{d} A \right)^H,
\label{eqie} \\
\bm C_{I\psi} & \equiv & \left(\int_{A_{pap}} \bs{\vec{\mathcal{\psi}}^e_{pap}} \textrm{d} A \right) \cdot
                  \left(\int_{A_{pap}} \bs{\vec{\mathcal{\psi}}^e_{pap}} \textrm{d} A \right)^H, 
\label{eqisi}
\eea
$k_B$ is the Boltzmann constant, $T_g$ is the ground temperature, $z_f$ is the free
space impedance, $S_{source}$ is the flux density of the observed source 
and $\bs{\vec{\mathcal{E}}^e_{pap}}$ and 
$\bs{\vec{\mathcal{\psi}}^e_{pap}}$ are the aperture fields (see Roshi \& Fisher 2016) due to 
the VEB, $\bs{\vec{\mathcal{E}}^e}$ and the CEB, $\bs{\vec{\mathcal{\psi}}^e}$ respectively. 
In Eqs.~\ref{eqce1} \& \ref{eqcsi1} the integration is over the
parts of the beam solid angle, $\Omega_{spill}$, seeing the ground radiation field and
in Eqs.~\ref{eqie} \& \ref{eqisi} the integration is over the aperture
plane, $A_{pap}$, of physical area $A_{ap}$.
The model equations
that are not affected by the embedded beam pattern definition are
\bea
\bm R_{rec} & = & 4 k_B T_0 \; \Big(R_n \bs{\mathcal{I}} + \sqrt{R_n g_n} \;\big(\rho \bm Z + \rho^* \bm Z^H\big) + g_n \bm Z\bm Z^H \Big), \\
T_n & =& T_{min} + N T_0 \; \frac{\bm w_1^H (\bm Z - Z_{opt} \bs{\mathcal{I}})
   (\bm Z - Z_{opt} \bs{\mathcal{I}})^H \bm w_1} {\mbox{Re}\{Z_{opt}\} \; \frac{1}{2}\bm w_1^H (\bm Z + \bm Z^H) \bm w_1}, \\
\bm R_{cmb} & =& 2 k_B T_{cmb} (\bm Z + \bm Z^H), \\
\bm R_{sky} & \approx& 2 k_B T_{sky} (\bm Z + \bm Z^H).
\eea
Here $\bm R_{rec}$, $\bm R_{cmb}$, $\bm R_{sky}$ are the open circuit voltage correlations
due to the amplifier noise, the cosmic microwave background and the sky background
radiation respectively,
$T_n$ is the receiver temperature of the PAF, $T_0 = 290 K$, $R_n$, $g_n$ and $\rho$ 
are the noise parameters of the amplifier,
which can equivalently be expressed in terms of the minimum noise temperature $T_{min}$,
Lange invariance $N$ and optimum impedance $Z_{opt}$ (Pospieszalski 2010); 
$T_{cmb}$ is the cosmic microwave
background temperature and $T_{sky} = T_{cmb} + T_{bg,\nu_0} \left(\frac{\nu}{\nu_0}\right)^{-2.7}$
is the temperature of the sky background at the observed off-source
position, $T_{bg,\nu_0}$ is the galactic background radiation
temperature at $\nu_0$, and $\nu$ is the frequency at which $\bm R_{sky}$
is computed.

\section{Embedded beam patterns from the CST far-field patterns}
\label{A8}

The CST (\verb|https://www.cst.com/|) microwave studio provides the
far-field pattern $\vec{E'}_j$ when the $j^{th}$ port is excited and all
other ports are terminated with the CST port impedance (in our case it is 50 $\Omega$).
From Eqs.~\ref{Evsvembpat} \& \ref{Evsiembpat} we get
\be
\vec{E'}_j = \sum_{i=1,M} q_{ij} \; \vec{E}^e_i,
\label{cstembed}
\ee
\be
\vec{E'}_j = \sum_{i=1,M} \mathcal{J}_{ij} \; \vec{\Psi}^e_i.
\label{cstembed1}
\ee
Here $q_{ij} = v_{0_i}$ is the port voltage and $\mathcal{J}_{ij}$ is the
port current. These voltages and currents are computed below. 
The elements of the wave amplitude vector for the excitation are 
\bea
a_i & =  & \sqrt{2\,P_{stim}} \quad \textrm{for} \; i=j \nonumber \\
    & =  & 0 \quad\quad\quad\quad\;\; \textrm{for}\; i \neq j
\eea
where $P_{stim} = 0.5$ W, is the RMS excitation power in the CST simulation. 
The wave amplitude vector $\bm b$ is then
\be
\bm b = a_j \begin{bmatrix}
    S_{1j} \\
    S_{2j} \\
    \threevdots \\
    S_{jj} \\
    \threevdots \\
    S_{Mj}
\end{bmatrix}
,
\ee
where $a_j$ is the $j^{th}$ element of the vector $\bm a$,
$S_{ij}, i = 1$ to $M$ is the $j^{th}$ column of $\bm S$. The
port voltages and currents are then
\bea
q_{ij} & = & \sqrt{z_0} (a_i + b_i) \nonumber \\
       & = & \sqrt{z_0} (1 + S_{jj}) a_j \quad \textrm{for}\;\;\; i = j \nonumber \\
       & = & \sqrt{z_0} S_{ij} a_j \quad\quad\quad \textrm{for}\;\;\; i \neq j \label{qij}\\
\mathcal{J}_{ij} & = & \frac{1}{\sqrt{z_0}} (a_i - b_i) \nonumber \\
       & = & \frac{1}{\sqrt{z_0}} (1 - S_{jj}) a_j \quad \textrm{for}\;\;\; i = j \nonumber \\
       & = & \frac{-1}{\sqrt{z_0}} S_{ij} a_j \quad\quad\quad \textrm{for}\;\;\; i \neq j
\label{jij}
\eea
The set of far-field patterns provided by the CST along with
the port voltages and currents can be used to obtain the VEB,
$\bs{\vec{E}^e}$ and the CEB, $\bs{\vec{\Psi}^e}$. 
Eq.~\ref{cstembed} \& \ref{cstembed1} for the set of far-field patterns
can be concisely written as
\bea
\bs{\vec{E}^{'}} & = & \bm Q \; \bs{\vec{E}^e}, \\
\bs{\vec{E}^{'}} & = & \bm J \; \bs{\vec{\Psi}^e}.
\eea
where the elements of the matrix $\bm Q$ are $q_{ij}$ and that of the 
matrix $\bm J$ are $\mathcal{J}_{ij}$.
This equation is valid for each $\theta, \phi$.
Using Eqs.~\ref{qij} \& \ref{jij} $\bm Q$ and $\bm J$ can be written as
\bea
\bm Q & = & \sqrt{2\, z_0\, P_{stim}} \;\;(\bs{\mathcal{I}} + \bm S),\\
\bm J & = & \sqrt{\frac{2\, P_{stim}}{z_0}} \;\;(\bs{\mathcal{I}} - \bm S).
\eea
The matrices $\bm Q$ and $\bm J$ are also related through the equation 
\be
\bm J = \bm Q \bm Z^{-1}.
\ee
The embedded beam patterns are then obtained as
\bea
\bs{\vec{E}^e} & = & \bm Q^{-1} \; \bs{\vec{E}^{'}}, \\
\bs{\vec{\Psi}^e} & = & \bm J^{-1} \; \bs{\vec{E}^{'}}.
\eea

\section{Some sanity checks}

\subsection{Energy conservation}
We verify here whether the computed embedded beam patterns 
satisfy energy conservation.
Details of such a verification for the VEB $\bs{\vec{E}^e}$ are given in
Roshi \& Fisher (2016).
We consider below the case for CEB, $\bs{\vec{\mathcal{\psi}}^e}$. 
From the definition of embedded beam pattern $\vec{\mathcal{\psi}}^e_j$  
the port currents are
\bea
\mathcal{J}_{0_i} & = & 1\; \textrm{A} \quad \textrm{for}\;\;\; i = j, \nonumber \\
    & = & 0\; \textrm{A} \quad \textrm{for}\;\;\; i \neq j,
\eea
and hence the wave amplitudes are 
\bea
\frac{1}{\sqrt{z_0}} (a_i - b_i) & = & 1 \quad \textrm{for}\;\;\; i = j, \nonumber \\
                                 & = & 0 \quad \textrm{for}\;\;\; i \neq j. 
\eea
The vector $\bm a$ can be written as
\be
\bm a = \bm b + \sqrt{z0} \begin{bmatrix}
    0 \\
    0 \\
    \threevdots \\
    1 \\
    \threevdots \\
    0
\end{bmatrix},
\ee
where the non-zero element (which is 1) is located at $j^{th}$ row. Substituting
this in the equation $\bm b = \bm S \bm a $ 
and re-arranging we get
\be
\bm b = \sqrt{z0}\;(\bs{\mathcal{I}} - \bm S)^{-1} \begin{bmatrix}
    S_{1j} \\
    S_{2j} \\
    \threevdots \\
    S_{jj} \\
    \threevdots \\
    S_{Mj}
\end{bmatrix}.
\ee
Power dissipated at the $j^{th}$ port is
\bea
P_{dis} & = & \frac{1}{2} (a_j a_j^* - b_j b_j^*), \\
        & = & \frac{\sqrt{z_0}}{2} (\sqrt{z_0} + (b_j + b_j^*)).
\label{pdis1}
\eea
The far-field beam pattern of the PAF for the above excitation is the
embedded beam pattern $\vec{\mathcal{\psi}}^e_j$ and hence
the radiated power is,  
\bea
P_{rad} & = & \frac{1}{2 z_f}\int_{sphere} \vec{\mathcal{\psi}}^e_j \cdot \vec{\mathcal{\psi}}^{e*}_j \; \textrm{d}A, \nonumber \\
        &  = & \frac{1}{2 z_f}\int_{4\pi} \vec{\Psi}^e_j  \cdot \vec{\Psi}^{e*}_j \; \textrm{d}\Omega.
\eea
For loss-less PAF $P_{dis} = P_{rad}$. This equality is satisfied in our
PAF model computation. Further, for a loss-less antenna,
\be
P_{rad} = \frac{1}{2} \mathcal{J}_{0_j}^2 \; \textrm{Re}\{Z_{pin_j}\},
\ee
where $\mathcal{J}_{0_j}$ is the current flowing to port $j$, which for the embedded pattern
$\vec{\mathcal{\psi}}^e_j$ is 1 A and $Z_{pin_j}$ is the input
impedance of port $j$ when all other ports are open circuited. The input
impedance for this case is given by
\be
Z_{pin_j} = z_{jj},
\label{embZin}
\ee
where $z_{jj}$ is the $j^{th}$ diagonal element of the impedance matrix ${\bm Z}$.
Thus
\be
P_{rad} = \frac{1}{2} \textrm{Re}\{z_{jj}\}.
\label{embpdis}
\ee

\subsection{PAF in a thermal radiation field}

In this Section, we show that the open circuit voltage correlations $\bm R_t$,
obtained from the two embedded beam patterns, when the the PAF is
embedded in a black body radiation field are equal to the result given by
Twiss's theorem (Twiss 1955). 
The correlation $\bm R_t$ is given by (Roshi \& Fisher 2016) 
\bea
\bm R_t & = &\frac{4 k_B T_0}{z_f} \bm Z \left(
\int_{4\pi} \bs{\vec{E}^e}\cdot \bs{\vec{E}^e}^H \textrm{d}\Omega \right) \bm Z^H, \nonumber \\
        & = & \frac{4 k_B T_0}{z_f} \bm Z \bm C_{Ce} \bm Z^H,
\label{thcorr1}
\eea
for the VEB, $\bs{\vec{\mathcal{E}}^e}$ and 
\be
\bm R_t  =  \frac{4 k_B T_0}{z_f} \bm C_{C\psi},
\label{thcorr2}
\ee
for the CEB, $\bs{\vec{\mathcal{\psi}}^e}$.
For a loss-less antenna the power dissipated at the ports should be equal to the radiated power,
which can be used to calculate $\bm C_{Ce}$ and $\bm C_{C\psi}$. The 
energy balance condition gives, 
\bea
\frac{1}{2} \left(\frac{\bm V_0^H \bm I_0}{2} + \frac{\bm I_0^H \bm V_0}{2}\right)  & = & \frac{1}{2 z_f} \bm V_0^H \bm C_{Ce} \bm V_0, \nonumber \\
\frac{1}{4} \bm V_0^H \left(\bm Z^{-1} + \left(\bm Z^{-1}\right)^H \right) \bm V_0 & = &
\frac{1}{2 z_f} \bm V_0^H \bm C_{Ce} \bm V_0,
\label{ce}
\eea
and
\bea
\frac{1}{2} \left(\frac{\bm V_0^H \bm I_0}{2} + \frac{\bm I_0^H \bm V_0}{2}\right)  & = & \frac{1}{2 z_f} \bm I_0^H \bm C_{C\psi} \bm I_0, \nonumber \\
\frac{1}{4} \bm I_0^H \left(\bm Z + \bm Z^H \right) \bm I_0 & = &
\frac{1}{2 z_f} \bm I_0^H \bm C_{C\psi} \bm I_0.
\label{csi}
\eea
Since Eqs.~\ref{ce} \& \ref{csi} are valid for arbitrary excitations it follows
that
\bea 
\frac{1}{2} \left(\bm Z^{-1} + \left(\bm Z^{-1}\right)^H \right) & = &
\frac{1}{z_f} \bm C_{Ce}, \label{impbc1} \\
\frac{1}{2} \left(\bm Z + \bm Z^H \right) & = &
\frac{1}{z_f} \bm C_{C\psi}. 
\label{impbc2}
\eea
Substituting Eq.~\ref{impbc1} in Eq.~\ref{thcorr1} and Eq.~\ref{impbc2} in 
Eq.~\ref{thcorr2}, we get 
\be
\bm R_t = 2 k_B T_0 \Big(\bm Z + \bm Z^H\Big),
\ee
from both Eqs.~\ref{thcorr1} \& \ref{thcorr2},
which is the voltage correlation given by Twiss's theorem (Twiss 1955). 

\section*{Acknowledgment}

I thank Rick Fisher and Bill Shillue for carefully proof reading the report and providing useful
comments.

\section*{References}

\noindent 
Pospieszalski, M. W., 2010, IEEE Microwave Magazine, 11, 61 

\noindent
Roshi, D. A., Fisher, J. R., 2016, NRAO, Electronics division internal report, 330. \\
\url{https://library.nrao.edu/public/memos/edir/EDIR_330.pdf}

\noindent
Twiss, R. Q., J. Appl. Phys., 1955, 26(5) 599.

\end{document}